\documentclass[reprint, amsmath,amssymb,aps, prl]{revtex4-2}

\usepackage{dcolumn}
\usepackage{bm}
\usepackage{braket}

\newcommand\identity{1\kern-0.25em\text{l}}

\usepackage{amsmath,amssymb,amsfonts} 
\usepackage{graphicx} 
\usepackage{hyperref}
\usepackage{bm} 
\usepackage{times} 
\usepackage{xcolor}
\usepackage{subcaption}
\usepackage{caption}
\DeclareCaptionStyle{mystyle}
  {format=plain,%
    textformat=period,
    justification=RaggedRight,
    singlelinecheck=true,
  }

\DeclareCaptionStyle{singlelinecentered}
  [justification=Centering]
  {style=mystyle}

\DeclareCaptionStyle{singlelineraggedleft}
  [justification=RaggedLeft]
  {style=mystyle}

\usepackage{mathrsfs}

\begin{document}

\preprint{APS/123-QED}

\title{Freezing and Shielding under Global Quenches for Long-Range Many-Body Systems}

 \author{Daniel Arrufat-Vicente}%
\affiliation{Institute for Theoretical Physics, ETH Z{\"u}rich, Wolfgang-Pauli-Str. 27, 8093 Z{\"u}rich, Switzerland}%
 \email{darrufat@phys.ethz.ch}
 \author{Nicol\`{o} Defenu}
 \affiliation{Institute for Theoretical Physics, ETH Z{\"u}rich, Wolfgang-Pauli-Str. 27, 8093 Z{\"u}rich, Switzerland}
\date{\today}

\begin{abstract}
We investigate how local perturbations spread on long-range interacting spin chains and show how their non-equilibrium behavior exactly matches the one of a Fermi-Hubbard model with long-range hoppings. We then focus on the thermalization of the fermionic model following a sudden quench of local interactions between two distinc fermionic species. For both models, the quasi-particle spectrum consists of gapped low-energy levels while the high-energy ones accumulate at a single point. When the low-energy gaps are large enough, turning on the local interactions does not effectively mix high and low-energy modes, resulting in a freezing of the dynamical evolution. In general, the states within the excited subspace are shielded from the long-range signatures of the model, which are responsible for the low-energy extended states, and the dynamics effectively look as if long-range couplings were absent. This shielding mechanism becomes evident in presence of disorder. It is shown that shielding and freezing are a universal feature of long-range interacting systems and their relation to the existence of quasi-stationary states is conjectured.
\end{abstract}

\maketitle

\emph{Introduction:} modern experimental quantum platforms allow to address the out-of-equilibrium dynamics and to tailor the interaction shape of quantum many-body  Hamiltonians~\cite{polkovnikov2011colloquium}. Out of this exciting possibility, a great interest has spurred in the physics of long-range interacting quantum systems\,\cite{defenu2023long}. Conventionally, the term long-range denotes couplings decaying as $1/r^\alpha$ (where $r$ is the distance between two components of the many-body system under study). Depending on the dimension of the system $d$, one defines three regimes: i) the local regime $\alpha>\alpha_{*}$, where $\alpha_{*}$ is a system and phenomenon dependent threshold beyond which nearest-neighbor or local physics is recovered; ii) the weak long-range regime ($d<\alpha<\alpha_{*}$) where long-range interactions are relevant, but thermodynamic functions remain extensive; iii) finally,  for $0<\alpha<d$ the strong long-range regime appears where conventional thermodynamics becomes ill-defined\,\cite{defenu2023outofequilibrium}. Long-range quantum systems count several practical realization including Rydberg atom arrays\,\cite{saffman2010quantum}, dipolar systems\,\cite{chomaz2023dipolar,carr2009cold}, trapped ion setups\,\cite{monroe2021programmable} and cold atoms in cavity experiments\,\cite{ritsch2013cold,mivehvar2021cavity,wu2023signatures} and are key candidates for the realization of efficient quantum computing devices\,\cite{ladd2010quantum}. 

In Ref.\,\cite{celardo2016shielding,santos2016cooperative} it was observed that, even in presence of strong long-range interactions, the dynamics were governed by an emergent short-range Hamiltonian, at least for initial states within specific subsets of the Hilbert space. 
In this case, the spreading of perturbation is confined to an effective light cone, as defined by the local \textit{Lieb-Robinson} bound~\cite{lieb1972finite} and it deviates from the general expectations for long-range interactions~\cite{feig2015nearly}. 
It is then said that such portion of the Hilbert space is \emph{shielded} from the effect of long-range interactions.

In the present work, we intend to provide solid evidences for the universality of the shielding phenomenon as well as to give further indications for the lack of thermalization produced by long-range interactions. Our approach is complementary to previous studies concerning two main aspects:
\begin{enumerate}
\item The shielding phenomenon is shown to be universal as it occurs on equal footing for all values of $\alpha$ and it can be traced back to the overall properties of the weakly interacting quasi-particle spectrum.
\item Our study also considers global quenches rather than only local ones\,\cite{celardo2016shielding,santos2016cooperative} enlarging the applicability of the concept of ``shielding". 
\end{enumerate}

In order to provide solid arguments to the above statements, we study the dynamics of a spin model where the interactions are strongly long-ranged as well as a Fermi-Hubbard model with long-ranged hopping amplitudes, both in $d=1$.\,\cite{hubbard1963electron}. On the one hand, the choice of these two models serves the scope of proving universality of the shielding phenomenon, which will appear in both models. On the other hand, the existence of a wide range of established tools to study the dynamics of Fermi-Hubbard models such as dynamical mean-field theory (DMFT)\,\cite{georges1996dynamical, rigol2008thermalization, tsuji2013nonthermal, aoki2014nonequilibrium} and unitary perturbation theory (UPT)\,\cite{moeckel2008interaction,kollar2011generalized} make such system an ideal playground for our study. 

As we will demonstrate, both models admit a well-defined quasi-particle limit in the weak-interaction regime, which allows us to justify our conclusions based on the spectral properties of low-energy quasi-particles. As shown below, it is the discreteness of the quasi-particle spectrum that drives the phenomena of shielding and freezing. Since long-range couplings generate discrete spectra for any $d > \alpha$, our results can be readily generalized to dimensions $d>1$. Furthermore, the detailed structure of the underlying lattice plays no significant role, as the long-range signatures are encoded exclusively in the low-momentum modes, while interactions among different fermionic species remain local~\cite{defenu2021metastability}.

We start by demonstrating that the density of states (DOS) of the Fermi-Hubbard model for a single fermion in the lattice is in exact correspondence to the DOS of an XX model with long-range interactions and a single spin flip on top of a magnetized background. We then provide numerical evidence on how this equivalence imprints the same dynamical behavior for more than one initially localized excitations in both systems. Finally, we consider the dynamics under global quenches. In this case, the large time dynamics can only be accessed for the Hubbard Hamiltonian, where UPT and DMFT calculation apply.

As an example of a global quench we consider the case where the local density-density interactions in between different fermionic species, initially absent, are suddenly ignited in the whole system. 
For purely long-range couplings the system's observables remain mostly unchanged after the quench, a phenomenon often referred to as \emph{freezing}. We analytically connect the existence of such phenomenon with the survival of low-energy single-particle spectral gaps and the appearance of an high-energy accumulation point in the quasi-particle spectrum.  Once disorder is introduced in the couplings, the degeneracy of the accumulation point is lifted, generating the celebrated semi-circular density of states (DOS)\,\cite{mehta2004random}. 

The appearance of the continuous part of the spectrum unfreezes the dynamics and the effect of long-range couplings remain confined to few low-energy discrete states, whose very limited impact on the global dynamics leads to the phenomenon of \emph{shielding}. These dynamical effects can be demonstrated exactly only in the thermodynamic limit, making them truly many-body phenomena. Finally, we discuss the differences between re-scaled and un-scaled long-range interactions under Kac's prescription and determine their effect in the dynamics of fermionic systems~\cite{kac2004onthevanderwaals}.

\emph{The models:} As discussed in the introduction, our study focuses on two long-range Hamiltonians, which share the same low-energy features. The spin-Hamiltonian describes two long-range interacting spin chains with local interchain exchange terms
\begin{subequations}
\begin{align}
    &H_0=-\sum_{i=1}^{L}\sum_{r=1}^{L/2-1}t_{r}\left(\sigma^+_{i}\sigma^{-}_{i+r} + \text{h.c}\right) - h\sum_{i}\sigma^z_{i}\label{eq:spin_chains}
    \\
    &\hspace{0.85cm}-\nonumber
    \sum_{i=1}^{L}\sum_{r=1}^{L/2-1}t_{r}\left(\bar{\sigma}^+_{i}\bar{\sigma}^-_{j+r} +\text{h.c}\right) - h\sum_{i}\bar{\sigma}^z_{i},
    \\
    &H_{\text{int}}=U\sum_{i=1}^{L}\sigma^z_i \bar{\sigma}^z_i \label{eq:spin_chains_int}
\end{align}
\end{subequations} 
where $\sigma^{\alpha}_i,\bar{\sigma}^{\alpha}_i$ are the Pauli matrices acting at site $i$ for the first and second chain respectively.
The second Hamiltonian describe a one dimensional Fermi-Hubbard model with two distinc fermionic species labeled by $\tau\in \{\uparrow,\downarrow\}$ decorated with long-range hopping amplitudes
\begin{subequations}
\begin{align}
    &H_0=-\sum_{\tau=\uparrow,\downarrow,i=1}^{L}\left(\sum_{r=1}^{L/2-1}t_r\Bigg(\hat{c}^{\dagger}_{i\tau}\hat{c}_{i+r\tau}+\text{h.c.}\right)\label{eq1a}
    \\
    &\hspace{5.5cm}+\mu\hat{c}^{\dagger}_{i\tau}\hat{c}_{i\tau}\Bigg)\nonumber ,\\
    &H_{\text{int}}=U\sum_{i=1}^{L}\hat{n}_{i\uparrow}\hat{n}_{i\downarrow}\label{eq1b},
\end{align}
\label{eq:fermi_hubbard}
\end{subequations}
In both cases the total Hamiltonian is $ H=H_{0}+H_\text{int}$. Under certain approximations, the two models can be connected via a Jordan-Wigner transformation~\footnote{See Sec.\,I of the supplementary material.}

The spins in Eq.~\eqref{eq:spin_chains} are coupled by an intrachain long-range exchange term with couplings $t_r$ and feel a homogeneous transverse magnetic field $h$. A local interchain transverse exchange coupling completes the analogy with the case of the Fermi-Hubbard Hamiltonian in Eq.\,\eqref{eq1a}.
Indeed, Hamiltonian~\eqref{eq:fermi_hubbard} features power-law decaying hopping amplitudes $t_r$ and a local density-density interaction term between the two fermionic species. At $U=0$ the fermionic Hamiltonian admits an analytic solution~\cite{lieb1961two}, while, for $U>0$ integrability is disrupted. As we argue in the following, both Hamiltonians in Eq.~\eqref{eq1a} and Eq.~\eqref{eq:spin_chains} are described by local-interacting quasi-particles with the non-analytic spectrum characteristic of long-range couplings. The latter description is a low-energy effective theory for the spin Hamiltonian~\eqref{eq:spin_chains} and applies only to quenches with low quasi-particle density.

In addition to the conventional long-range envelope, we allow the exchange couplings $t_r$ to have  Gaussian fluctuations around a mean-value that decays as a power-law of the distance. 
\begin{align}\label{eq:gaussian_prob}
    \braket{t_r} = \frac{M_\alpha}{r^\alpha}, \hspace{1cm} \braket{t_rt_{r'}}=\sigma^2\delta_{rr'}\equiv \frac{J^2}{L}\delta_{rr'}\,.
\end{align}
For the spin Hamiltonian in Eq.\,\eqref{eq:spin_chains} a rescaling of the long-range exchange term is needed to maintain the internal energy of the system extensive, a procedure refereed to as Kac's re-scaling\,\cite{defenu2023long}.
The same re-scaling of the hopping amplitudes in the long-range fermionic model in Eq.\,\eqref{eq1a} will instead result in a subextensive kinetic energy~\footnote{As an example, if the hoppings are purely non-local ${(\alpha=0)}$ the kinetic energy is solely carried by the $k=0$ momentum mode~\cite{vandongen1989exact}}. 
However, in order to draw connections with most of the existing literature, we will consider both Kac-scaled and un-scaled strong long-range interactions.

\emph{Density of states}: let us first consider Hamiltonian~\eqref{eq1a} with $\alpha=0$. There, the quasi-particle spectrum features a $L-1$ degenerate state with energy $\varepsilon_{k>0}=\mu$, which we refer to as the high-energy accumulation point~\cite{dellanna2008critical}. A finite disorder strength $\sigma>0$ lifts the degeneracy and leads to a continuous DOS with semi-circular shape plus the low-energy isolated state at $k=0$, in analogy to the problem of a single impurity in a crystal~\cite{edwards1976theigenvalue}. Similarly, for $0\leq\alpha<1$ in absence of disorder the low-energy part of spectrum remains discrete while the high energy modes accumulate near a single point in the thermodynamic limit. The presence of disorder mainly mixes the accumulated high energy modes while keeping the low-energy ones gapped, the DOS takes the generic form valid at any $\alpha<1$~\footnote{See Sec.\,II of the supplementary material},
\begin{equation}\label{eq:DOS}
    \rho(\lambda)=\frac{\sqrt{4J^2-\lambda^2}}{2\pi J^2}+\frac{1}{L}\sum_{n<n^*}\delta(\lambda-\lambda_n)
\end{equation}
where $\lambda_n=M_\alpha\varepsilon_n+J^2/(M_\alpha\varepsilon_n)$, $\varepsilon_n$ corresponds to the eigen-energies of the free long-range hopping model ($\alpha<1$) in the absence of disorder 
and $n^{*}$ is a critical $n$ for which: 
$|\lambda_{n^*}|\leq 2J$ and $|\lambda_{n^*-1}|> 2J$ where $n \in \mathbb{Z}$ and ${-\frac{L}{2} \leq n < \frac{L}{2}}$. 
The eigen-energies in absence of disorder $\varepsilon_n$ read
\begin{subequations}\label{eq:dispersion}
\begin{align}
    \varepsilon_n=&-L^{-1}_{\alpha}\sum_{r=1}^{L/2}\frac{\cos(2\pi n r/L)}{r^{\alpha}}
\approx-c_\alpha\int_0^{1/2} ds\frac{\cos\left(2\pi n s\right)}{s^{\alpha}}
\end{align}
\end{subequations}
with $L^{-1}_{\alpha}=c_{\alpha}L^{\alpha-1}=(1-\alpha)2^{(1-\alpha)}L^{\alpha-1}$~\cite{defenu2021metastability}. The latter equality in Eq.\,\eqref{eq:dispersion} is strictly valid in the thermodynamic limit ${L\to \infty}$ and the omission of Kac's renormalization $L_\alpha$ leads to eigen-energies increasing with the volume $L$
\begin{align}
\varepsilon_{n}\propto\begin{cases}
 O(1)&\qquad \mathrm{Kac\,\,on}.\\
O(L^{1-\alpha})&\qquad \mathrm{Kac\,\,off} .
\end{cases}
\end{align}
It is worth reminding that in Eq.\,\eqref{eq:DOS} the summation over discrete energy states only includes the $n^{*}$ 
eigenvalues whose energy lies outside the continuous portion of the spectrum, which, for Kac-scaled interactions, remains finite in the thermodynamic limit.
\begin{figure}
    \centering
\includegraphics[width=.99\linewidth,trim={0 0 0 0},clip]{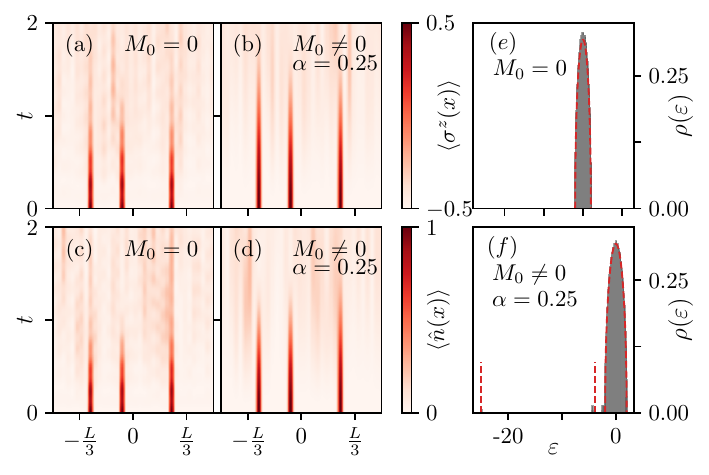}
    \caption{Panels~(a)-(b): Evolution of the magnetization profile under $H_0$ from Eq.~\eqref{eq:spin_chains} for three initially localized perturbations (spin flips on top of a magnetized background). Panel (a) corresponds to the dynamics unders purely off-diagonal disorder ($M_0=0$) while panel (b) in addition to disorder contains the long-range envelope with $M_0 = -4\pi$ and $\alpha=0.25$. Panel~(c)-(d): Dynamics for three initially localized fermions under $H_0$ from Eq.~\eqref{eq1a} under the same parameters as panels (a) and (b) respectively.  Both calculations were performed with $L=36$. Panels~(e)-(f): Density of states (DOS) of a single-particle excitation. Panel (e) shows the DOS with Gaussian disorder, as specified in Eq.\,\eqref{eq:gaussian_prob}. Panel (f) includes additional long-range hoppings with $M_0 = -4\pi$ and $\alpha=0.25$. In both panels, the shaded area corresponds to a normalized eigenvalue histogram for $L=2^{13}$, while the red line sketches the theoretical prediction given by Eq.~\eqref{eq:DOS} with the dashed line corresponding to the continuous predictions and the tiny red columns highlight the discrete eigenvalues}
    \label{Fig1}
\end{figure}

Without disorder, the spin Hamiltonian~\eqref{eq:spin_chains} at $\alpha=0$ corresponds to the fully-connected XX model. There, the many-body spectrum can be computed exactly. The Hamiltonian eigenstates ($|S,S_z\rangle$), correspond to the eigenstates of the total spin $S$ and total magnetization $S_z$~\cite{lipkin1965validity}. Assuming $L$ to be even and considering the total magnetization to be fixed to $S_z = (L/2-1)$, the spectrum consists of $L-1$ energetically degenerate states and a single gapped state with the energy gap given by $\Delta\varepsilon=M_{0}$. The unique gaped state corresponds to the state with maximal total spin $S=L/2$, while the  $L-1$ degenerate eigenstate are composed by the different $S\in \{0,\cdots,L/2-1\}$ eigenstates. Such degeneracies come from the different possible ways
to arrange the microscopic $1/2$-spins in order to form a
total spin $S$~\cite{delre2016nonequilibrium}. In analogy to the fermionic model, disorder primarily affects the $S=L/2-1$ eigenstates altering their energy and lifting the degeneracy,  but leaves the isolated $S=L/2$ state almost unaltered. 

Although formulated for the case $\alpha = 0$, the analogy between the quasi-particle spectra of the two models extends to arbitrary values of $\alpha$ in the low-energy region of the spectrum (see Fig.~\ref{Fig1}(e–f) and~\footnote{See supplemental material}). Consequently, it is natural to expect that any low-energy state with sufficiently low quasi-particle density will exhibit similar dynamical behavior under Hamiltonians~\eqref{eq:spin_chains} and~\eqref{eq1a}.  Fig.~\ref{Fig1}~(a) illustrates the dynamics of Hamiltonian~\eqref{eq:spin_chains} in the presence of disorder but without long-range interactions ($M_{\alpha}=0$), starting from a state fully polarized along the magnetic field except for three isolated spin flips. Panel (b) shows the dynamics when long-range interactions are included ($M_{\alpha}>0$, $\alpha\simeq 1/4$).
The correspondence between spin dynamics and fermionic dynamics becomes evident when comparing Fig.~\ref{Fig1}(a,b) with Fig.~\ref{Fig1}(c,d), which present the evolution under Hamiltonian~\eqref{eq:fermi_hubbard}. This similarity was anticipated from the comparison of the low-energy density of states (DOS) of the two models, see Fig.~\ref{Fig1}(e–f).

As we demonstrate in the following, the discreteness of the long-range quasi-particle spectrum lies at the root of the shielding phenomenon. Our calculations will be performed for the Hamiltonian~\eqref{eq:fermi_hubbard}, where, owing to the applicability of DMFT and UPT, we are going to prove that long-range interactions are shielded for any initial state. Due to the correspondence between the quasi-particle dynamics of the Fermi-Hubbard and XX Hamiltonians, our conclusions will also apply to the Hamiltonian~\eqref{eq:spin_chains}, at least for initial states in the low-energy regime considered also in Ref.~\cite{celardo2016shielding}. 

\emph{Freezing and its relation with the accumulation point}:
Shielding and freezing are two sides of the same coin. In the case of \emph{shielding}, the dynamical evolution of observables is unaffected by the presence of long-range couplings. Conversely, \emph{freezing} refers to observables remaining mostly unchanged after a quantum quench. Both phenomena occur in the strong long-range regime, so we focus on $\alpha < d$: Initially, the system is prepared in the non-interacting ground-state, i.e. the lowest energy state of the Hamiltonian $H_0$. 
The chemical potential is therefore tuned to ensure we are at half filling, $N_{\uparrow}=N_{\downarrow}=L/2$ where $N_{\uparrow(\downarrow)}$ is the number of the distinct electrons species. 
This corresponds to a filled Fermi sea up to the middle of the semi-circle in Eq.\,\eqref{eq:DOS}.

\begin{figure}[h!]
    \centering
    \includegraphics[width=.85\linewidth,trim={0.2cm 0.2cm 0.2cm 0.02cm},clip]{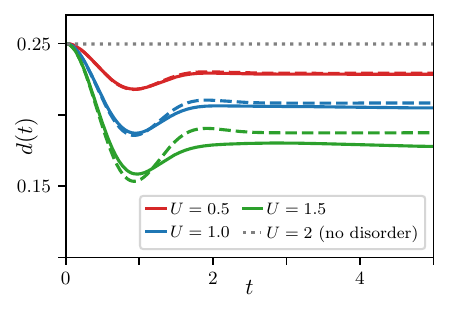}
      \caption{Double occupancy after a sudden quench. Continuous and  dashed lines correspond, respectively, to DMFT and UPT computations for the disordered case in absence of long-range tails ($M_\alpha=0$) for different on-site interaction $U=0.5,1.0,1.5$ while the dotted line is a DMFT computation of the clean fully connected Fermi-Hubbard model ($\alpha=0$) at $U=2$. We observe the freezing of dynamics in absence of disorder and a relaxation to an equilibrium value, which decreases as $U$ increases, in the presence of it\,\cite{eckstein2010interaction}}
    \label{Fig2}
\end{figure}

At $t=0$ the onsite interactions are turned on and the non-interacting ground-state evolved according to $H$~\eqref{eq:fermi_hubbard}. 
The result of this procedure is reported in Fig.\,\ref{Fig2} for $\alpha=0$. 
For the clean long-range system (dotted line) the double occupancy, $d\equiv \sum_{i}\braket{\hat{n}_{i\uparrow}\hat{n}_{i\downarrow}}$, 
remains fixed to its initial value after the interaction quench, so the electrons effectively behave as if they where not allow to hop in between sites in agreement with the expected suppression of excitation propagation evidenced 
for local quenches in long-range systems\,\cite{santos2016cooperative,celardo2016shielding}. 

For $\alpha<d$, the absence of dynamical signatures after turning on the interaction can be justified analytically within the 
DMFT formalism~\footnote{see Sec.\,III of the supplementary material}. 
The idea behind DMFT is to simplify the many-body problem by approximating it with to the problem of a 
single impurity coupled to a  bath. 
Therefore, the single electron Green function $G(i\omega_{n})$ looses its 
momentum dependence and depends only on the imaginary frequency. 
In order to obtain the impurity Green function one has to solve a self-consistent 
equation for the impurity self-energy $\Sigma(i\omega_{n})=\mathscr{G}_{0}^{-1}(i\omega_{n})-G^{-1}(i\omega_{n})$, 
where $\mathscr{G}_{0}(i\omega_{n})$ is the bare Green function of the impurity, also referred to as Weiss effective field~\cite{georges1996dynamical}.

As the thermodynamic limit is approached, the kintetic energy contribution from long-range couplings with $\alpha<d$ vanishes, due to
the kinetic energy contribution being dominated by the accumulation point. Thus, the impurity Green's function $\mathscr{G}_0(i\omega_{n})$ converges to the atomic limit~\footnote{See Sec.\,III of the supplementary material.}
\begin{equation}
\label{dmft_green_function}\mathscr{G}^{-1}_0(i\omega_n)=i\omega_n+\mu+\mathcal{O}(1/L)\,.
\end{equation}
The many-body/\emph{cooperative} nature of the shielding effect is evident by the requirement of a large number of particles to ensure that the $O(L^{-1})$ in 
Eq.~\eqref{dmft_green_function} is negligible and, thus, the kinetic energy effect vanish.
In fact, in the thermodynamic limit the kinetic and interaction terms effectively 
commute 
leading to the simultaneous conservation of kinetic and potential energy~\cite{vandongen1989exact}.
At $t\geq 0$, as the local density-density interactions in Eq.~\eqref{eq1b} are ignited, a finite potential energy is suddenly imprinted onto the system 
which cannot be dispersed in any way. 

In other words, the energy gaps between quasi-particle states remain finite even at large $L$ due to the strong long-range nature of the tunneling. This suppresses the proliferation of fluctuations following the quench, thereby preventing mode dephasing, which is fundamental to equilibration in many-body systems~\cite{lasinio1996chaotic}. As the thermodynamic limit is approached ($L\to\infty$), the dominant quasi-particle contribution to the dynamics becomes increasingly concentrated in the highest energy region of the spectrum—specifically, near the accumulation point. This leads to the erasure of all dynamical features, as illustrated by the dotted line in Fig.~\ref{Fig2}. A similar reasoning applies to spatial fluctuations, which are also suppressed by the strong long-range tunneling term~\cite{defenu2023long}. This justifies the applicability of DMFT even in the one-dimensional case.

\emph{From freezing to shielding:}
As previously argued, long-range couplings suppress spatial fluctuations, guaranteeing the applicability of mean-field approaches. 
In a similar way, Gaussian disorder ensures that momentum and energy conservation do not generate divergences 
in the unitary perturbation theory treatment of the problem. 
Thus, at half-filling, both in the DMFT and UPT study of the disordered case, 
turning on local density-density interactions at $t>0$ only 
triggers scattering processes that 
involve transitions between states within the continuous semi-circular portion 
of the spectrum. Therefore, the dynamics are shielded from long-range couplings, which only affect the discrete low-energy states in the DOS, see Fig.~\ref{Fig1}.

Indeed, the continuous semicircular portion of the spectrum in Eq.\,\eqref{eq:DOS} induced by disorder corresponds to the DOS of the local system in the Bethe lattice.
By comparing the dynamics in the presence and absence of disorder (see Fig.\,\ref{Fig2}) we can conclude how disorder is necessary in order to 
break the \emph{freezing} mechanism and have dynamical signatures.
The addition of the long-range coupling amplitudes in Eq.\,\eqref{eq:DOS} causes the appearance of a discrete set of low-energy quasi-particle states. However, these discrete states become statistically irrelevant in the thermodynamic limit, which removes the effect of long-range couplings. This is readily shown by the self-consistent condition between the Weiss function and the single electron Green's function within DMFT ${\mathscr{G}^{-1}_0(i\omega_n)=i\omega_n+\mu-J^2G(i\omega_n)+\mathcal{O}(1/L)}$ for the DOS in Eq.~\eqref{eq:DOS}~\footnote{See Supplemental Material Sec.~III.B.}. 

Similarly to the case of local quenches in the long-range spin models discussed in Fig.~\ref{Fig1} and Refs.~\cite{celardo2016shielding,santos2016cooperative}, the dynamics induced by Hamiltonian~\eqref{eq:fermi_hubbard} is not influenced by the discrete low energy states and its thermalization timescales are solely determined by energies near the Fermi-surface. The thermodynamic limit procedure is essential to obtain exact shielding of long-range couplings. Thus, in order to investigate the finite size signatures of shielding we employ the UPT approach, which has already been successfully applied to the local case\,\cite{hackl2008aunitary}. In UPT, the time evolution of an observable $\hat{A}$ can be computed by the expression

\begin{equation}
    \braket{A}_t=\braket{A}_0+4U^2\int_{-\infty}^{\infty}d\omega\frac{\sin^2\left(\omega t/2\right)}{\omega^2}J(\omega)+\mathcal{O}(U^3),
\end{equation}
where $J(\omega)=\braket{\hat{H}_{\text{int}}\left(\hat{A}-\braket{A}_0\right)\delta\left(\hat{H}_0-\braket{H_0}-\omega\right)\hat{H}_{\text{int}}}_0$ captures scattering processes up to second order in $U$.
\begin{figure}[t!]
    \centering
   \includegraphics[width=1\linewidth,trim={0cm 0.11cm 0cm 0cm},clip]{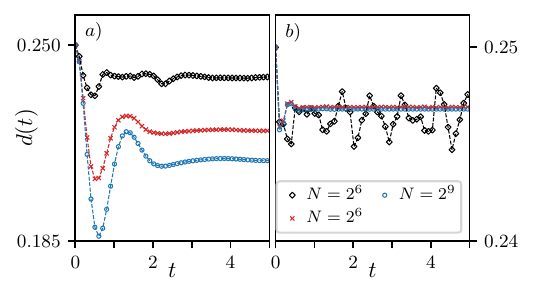}
    \caption{Double occupancy after a sudden quench to $U =1$ for $\alpha=0.5$, $\sigma=1$ and $M_0=-4\pi$. In panel a) the couplings have been re-escaled using Kac's prescription while for panel b) the couplings are un-scaled. Three different systems sizes are shown $L=2^{3},2^{6},2^{9}$ (black diamonds, red crosses and blue dots) from top to bottom close to $t=0$. For the smallest size $L=2^3$ we observe the presence of Poincar\'{e} recurrences which already dephase both for $L=2^{6},2^{9}$}
    \label{fig:UPT_DISORDER}
\end{figure}
In the local case $\alpha\to\infty$ UPT predicts a prethermalization plateau and the appearance of quasi-conserved charges, which lead to a generalized Gibbs distribution\,\cite{kollar2011generalized}. 

In Fig.\,\ref{fig:UPT_DISORDER} (left panel) we present the results for the  $\alpha=0.5$ case at half filling for different systems size $L=2^{3},2^{6},2^{9}$ both for scaled and unscaled couplings. By comparing the blue line on the bottom of Fig.~\ref{fig:UPT_DISORDER} (a) with the blue dashed line in the middle of Fig.~\ref{Fig2}, we observe how, as long as the system is considerably big, we recover the qualitative behavior of the infinite dimensional Bethe-Lattice~\cite{eckstein2010interaction}.
At small sizes, the spectrum can be considered discrete everywhere and Poincar\'{e} revivals are observed, black diamonds in Fig.~\ref{fig:UPT_DISORDER}. 
The presence or absence of the Kac's renormalization factor influences the magnitude on the relative change of the double occupancy before and after the quench, without adding any temporal structure, as we can see comparing the Kac scaled result (Fig.\,\ref{fig:UPT_DISORDER}a) with the unscaled one (Fig.\,\ref{fig:UPT_DISORDER}b). The quantitative discrepancy between the Kac scaled and unscaled cases originates from the change in spectral weight outside the semicircular DOS due to the Kac's factor.

\emph{Conclusions and perspectives:} Our analysis demonstrates that the dynamical evolution following local quenches of Fermi-Hubbard models with long-range hoppings mimics the one of spin chains with long-range exchange interactions, due to the correspondence between their quasi-particle spectra. The key characteristic of this dynamical evolution is \emph{shielding}: in the presence of off-diagonal disorder, local perturbations spread rapidly in both models, and at the same time, both systems become insensitive to the inclusion of long-range couplings.

In the Fermionic model shielding is demonstrated to extend also to global quenches, where a repulsive on-site interactions is suddenly ignited in the system. In the absence of disorder, electrons localize, a phenomenon which we refer to as \emph{freezing}. Disorder unfreezes the dynamics, but the system is shown to remain insensitive to additional long-range couplings, leading back to the phenomenology of shielding. We conclude that shielding is fundamentally linked to the low-energy (universal) spectral features of systems with long-range couplings. 

The correspondence between the low-energy spectra of Hamiltonians~\eqref{eq:spin_chains} and~\eqref{eq:fermi_hubbard} has been demonstrated exactly in the supplemental material, but our picture for local quenches applies to any model whose quasi-particle spectrum displays coexistence of continuous and discrete regions akin to the one described in Fig.~\ref{Fig1}, including the model studied in Refs.~\cite{celardo2016shielding,santos2016cooperative}. 

For global quenches, the appearance of freezing and shielding is peculiar to Fermi quasi-particles, where the effects of interactions are suppressed by the combined effects of statistics and spectral discreteness. For systems with Bose quasi-particles, such as long-range quantum spin Hamiltonians spectral discreteness appears to affect only specific states, i.e. the quantum scars~\cite{comparin2022robust}, while most of the spectrum becomes chaotic~\cite{lerose2023theory}. A clearer picture of the many‑body spectrum of long‑range spin Hamiltonians in the finite‑energy regime may be obtained by developing a quantum kinetic theory of long‑range models.

\emph{Acknowledgements:}
This research was supported in part by grant NSF PHY-1748958 to the Kavli Institute for Theoretical Physics (KITP). This research was funded by the Swiss National Science Foundation (SNSF) grant number 200021 207537, by the Deutsche Forschungsgemeinschaft (DFG, German Research Foundation) under Germany's Excellence Strategy EXC2181/1-390900948 (the Heidelberg STRUCTURES Excellence Cluster) and the Swiss State Secretariat for Education, Research and Innovation (SERI).

\onecolumngrid 
\vspace{1em}

%
%
\section*{Supplemental Material}
\section{Jordan Wigner Transformation.}
\onecolumngrid 
\emph{Long-range spin chains:}
In this section we show the connection between the fermionic model and the 
long-range spin chain. Recall the Hamiltonian of a single spin chain 
\begin{align}\label{eq:end1}
    \hat{H}=-\sum_{l<j}V_{|l-j|}\left(\hat{\sigma}^+_l\hat{\sigma}^-_j+\hat{\sigma}^-_l\hat{\sigma}^+_j\right)-h\sum_j\hat{\sigma}^z_j.
    \end{align}
We can map spins to fermions by using the Jordan-Wigner (JW) transformation
\begin{align}
\hat{\sigma}_j^z&=1-2\hat{c}_j^\dagger\hat{c}_j,\\
\hat{\sigma}_j^y&=-\mathrm{i}\Big[\prod_{m=1}^{j-1}\big(1-2\hat{c}_m^\dagger\hat{c}_m\big)\Big]\big(\hat{c}_j-\hat{c}_j^\dagger\big),\\
\hat{\sigma}_j^x&=-\Big[\prod_{m=1}^{j-1}\big(1-2\hat{c}_m^\dagger\hat{c}_m\big)\Big]\big(\hat{c}_j+\hat{c}_j^\dagger\big),
\end{align}
where $\hat{c}_j,\hat{c}_j^\dagger$ are fermionic annihilation and creation operators, respectively, which satisfy the anticommutation relations $\{\hat{c}_l,\hat{c}_j\}=0$ and $\{\hat{c}_l,\hat{c}_j^\dagger\}=\delta_{l,j}$. Plugging this relation onto Eq.~\eqref{eq:end1} one obtains a fermionic model with long-range hoppings
\begin{align}\label{eq:pauli_string}
    \hat{H}=&\,-2\sum_{l<j}V_{|l-j|}\Bigg\{\hat{c}_l^\dagger\Big[\prod_{n=l+1}^{j-1}\big(1-2\hat{c}_n^\dagger\hat{c}_n\big)\Big]\hat{c}_j
    -\hat{c}_l\Big[\prod_{n=l+1}^{j-1}\big(1-2\hat{c}_n^\dagger\hat{c}_n\big)\Big]\hat{c}_j^\dagger
    \Bigg\}-h\sum_j\big(1-2\hat{c}_j^\dagger\hat{c}_j\big).
    \end{align}
Hamiltonian~\eqref{eq:pauli_string} cannot be solved exactly due to the presence of terms of higher than quadratic order.
Nevertheless, close to a ferromagnetic state we can approximate
\begin{align}\label{eq:approx}
\prod_{n=l+1}^{j-1}\big(1-2\hat{c}_n^\dagger\hat{c}_n\big)\approx 1,
\end{align}
for every $j\geq l+2$, which is referred to as the \textit{truncated} JW transformation to obtain a quadratic form of the Hamiltonian
\begin{align}
\hat{H}=&\,-2 \sum_{l<j} V_{|l-j|}\big(\hat{c}_l^\dagger\hat{c}_j-\hat{c}_l\hat{c}_j^\dagger\big)-h\sum_j\big(1-2\hat{c}_j^\dagger\hat{c}_j\big).
\label{lrk_h}
\end{align}

\emph{Exact equivalence:}
The approximation in Eq.~\eqref{eq:approx} can be rigorously justified in the low-energy limit. Indeed, the Hamiltonian in the manuscript preserves the total magnetization $S_z=\sum_{i}\sigma^z_{i}/2$ in analogy with the conservation of fermions in the Fermi-Hubbard model of Eq.\,\eqref{lrk_h}. For state containing just a single excitation, i.e. a spin-flip, the Hamiltonian in Eq.~\eqref{eq:pauli_string} can be projected to the single quasi-particle sector
\begin{align}\label{eq:proj1}
    P\hat{H}P=&\,-2\sum_{l<j}V_{|l-j|}\Bigg\{P\hat{c}_l^\dagger\left(Q+P\right) \Big[\prod_{n=l+1}^{j-1}\big(1-2\hat{c}_n^\dagger\hat{c}_n\big)\Big]\left(Q+P\right)\hat{c}_j P
    \\
    &\nonumber -P\hat{c}_l\left(Q+P\right)\Big[\prod_{n=l+1}^{j-1}\big(1-2\hat{c}_n^\dagger\hat{c}_n\big)\Big]\left(Q+P\right)\hat{c}_j^\dagger P
    \Bigg\}-h\sum_j P\big(1-2\hat{c}_j^\dagger\hat{c}_j\big)P\,,
    \end{align}
where the projector $P$ and $Q$ are
\begin{align}
    &P= \sum_{l}\hat{c}^{\dagger}_{l}|0\rangle\langle0|\hat{c}_{l} \,,
    \\
    & Q= \identity - P = |0\rangle \langle 0 | + \frac{1}{2!}\sum_{l,j}\hat{c}^{\dagger}_{j}\hat{c}^{\dagger}_{l}
    |0\rangle\langle0|\hat{c}_{l}\hat{c}_{j} + 
\cdots \,.
\end{align}
Note how $Q$ is the orthogonal projection that projects onto subspaces with either more than one or zero quasi-particles. Hamiltonian\eqref{eq:proj1} then transforms to
\begin{align}\label{eq:proj2}
    P\hat{H}P=&\,-2\sum_{l<j}V_{|l-j|}\Bigg\{P\hat{c}_l^\dagger Q\Big[\prod_{n=l+1}^{j-1}\big(1-2\hat{c}_n^\dagger\hat{c}_n\big)\Big]Q\hat{c}_j P
-P\hat{c}_l Q\Big[\prod_{n=l+1}^{j-1}\big(1-2\hat{c}_n^\dagger\hat{c}_n\big)\Big] Q \hat{c}_j^\dagger P
    \Bigg\}\\\nonumber
    &-h\sum_j P\big(1-2\hat{c}_j^\dagger\hat{c}_j\big)P.
    \end{align}
where we made use of
\begin{align}
    &P\Big[\prod_{n=l+1}^{j-1}\big(1-2\hat{c}_n^\dagger\hat{c}_n\big)\Big]Q = 0 \,,
    \\
    &P\big(\hat{c}_j^\dagger\pm \hat{c}_j\big)P=0\,.
\end{align}
The non-conserving terms in Eq.\,\eqref{eq:proj1} at most can mix in between states with zero, one or two fermions so we can truncate the projector $Q$ to
\begin{equation}
    Q = |0\rangle \langle 0 | + \frac{1}{2!}\sum_{l,j}\hat{c}^{\dagger}_{j}\hat{c}^{\dagger}_{l}
    |0\rangle\langle0|\hat{c}_{l}\hat{c}_{j} \,,
\end{equation}
such that the projected Hamiltonian simplifies to
\begin{align}\label{eq:proj3}
P\hat{H}P=&\,-2\sum_{l<j}V_{|l-j|}\Bigg\{P \hat{c}_l^\dagger|0\rangle\langle 0 |\Big[\prod_{n=l+1}^{j-1}\big(1-2\hat{c}_n^\dagger\hat{c}_n\big)\Big]|0\rangle\langle 0 |\hat{c}_j P -P \hat{c}_l |0\rangle\langle 0 |\Big[\prod_{n=l+1}^{j-1}\big(1-2\hat{c}_n^\dagger\hat{c}_n\big)\Big]|0\rangle\langle 0 |\hat{c}_j^\dagger
\Bigg\}\\ \nonumber
&-2\sum_{l<j}V_{|l-j|}\Bigg\{P \hat{c}_l^\dagger\frac{1}{2!}\sum_{l_1,j_1}\hat{c}^{\dagger}_{j_1}\hat{c}^{\dagger}_{l_1}
    |0\rangle\langle0|\hat{c}_{l_1}\hat{c}_{j_1}\Big[\prod_{n=l+1}^{j-1}\big(1-2\hat{c}_n^\dagger\hat{c}_n\big)\Big]\frac{1}{2!}\sum_{l_2,j_2}\hat{c}^{\dagger}_{j_2}\hat{c}^{\dagger}_{l_2}
    |0\rangle\langle0|\hat{c}_{l_2}\hat{c}_{j_2} \hat{c}_jP
\\ \nonumber
&-P \hat{c}_l\frac{1}{2!}\sum_{l_1,j_1}\hat{c}^{\dagger}_{j_1}\hat{c}^{\dagger}_{l_1}
    |0\rangle\langle0|\hat{c}_{l_1}\hat{c}_{j_1}\Big[\prod_{n=l+1}^{j-1}\big(1-2\hat{c}_n^\dagger\hat{c}_n\big)\Big]|\frac{1}{2!}\sum_{l_2,j_2}\hat{c}^{\dagger}_{j_2}\hat{c}^{\dagger}_{l_2}
    |0\rangle\langle0|\hat{c}_{l_2}\hat{c}_{j_2}\hat{c}_j^\dagger P
\Bigg\}\\ \nonumber
&-h\sum_jP\big(1- 2\hat{c}_j^\dagger\hat{c}_j\big)P\,,
\end{align}
as the Pauli strings operators $\prod_{n=l+1}^{j-1}\big(1-2\hat{c}_n^\dagger\hat{c}_n\big)$ involve creation and annhilation of fermions in lattices in between the sites $l,j$ in the single fermion subspace only the product of identities term of the Pauli string survives.
\begin{align}\label{eq:proj4}
P\hat{H}P=&\,-2\sum_{l<j}V_{|l-j|}\Bigg\{P\hat{c}_l^\dagger|0\rangle\langle 0 |\hat{c}_j P
\Bigg\}-2\sum_{l<j}V_{|l-j|}\Bigg\{P\hat{c}_l^\dagger\frac{1}{2!}\sum_{l_1,j_1}\hat{c}^{\dagger}_{j_1}\hat{c}^{\dagger}_{l_1}
    |0\rangle\langle0|\hat{c}_{l_1}\hat{c}_{j_1}\hat{c}_j P
    \\ \nonumber
    &\qquad-P \hat{c}_l \frac{1}{2!}\sum_{l_1,j_1}\hat{c}^{\dagger}_{j_1}\hat{c}^{\dagger}_{l_1}
    |0\rangle\langle0|\hat{c}_{l_1}\hat{c}_{j_1}\hat{c}_j^\dagger P
\Bigg\}
-h\sum_jP\big(1- 2\hat{c}_j^\dagger\hat{c}_j\big)P.
\end{align}
Let us note the following properties
\begin{align}
    P \hat{c}_j^\dagger \frac{1}{2!}\sum_{l_1,j_1}\hat{c}^{\dagger}_{j_1}\hat{c}^{\dagger}_{l_1}
    |0\rangle\langle0|\hat{c}_{l_1}\hat{c}_{j_1} =& 0\,,
    \\
    P \hat{c}_j \frac{1}{2!}\sum_{l_1,j_1}\hat{c}^{\dagger}_{j_1}\hat{c}^{\dagger}_{l_1}
    |0\rangle\langle0|\hat{c}_{l_1}\hat{c}_{j_1} =& P \hat{c}_{j} \,,
    \\
    \frac{1}{2!}\sum_{l_1,j_1}\hat{c}^{\dagger}_{j_1}\hat{c}^{\dagger}_{l_1}
    |0\rangle\langle0|\hat{c}_{l_1}\hat{c}_{j_1} \hat{c}_j^\dagger P =& \hat{c}^{\dagger}_jP\,,
    \\
    \frac{1}{2!}\sum_{l_1,j_1}\hat{c}^{\dagger}_{j_1}\hat{c}^{\dagger}_{l_1}
    |0\rangle\langle0|\hat{c}_{l_1}\hat{c}_{j_1} \hat{c}_j P =& 0 \,,
\end{align}
which, when plugged into Eq.\eqref{eq:proj4}, finishes the proof
\begin{align}\label{eq:proj5}
P\hat{H}P=&\,-2\sum_{l<j}V_{|l-j|}\Bigg\{P\hat{c}_l^\dagger|0\rangle\langle 0 |\hat{c}_j P
\Bigg\}
 +2\sum_{l<j}V_{|l-j|}\Bigg\{P\big(\hat{c}_l \hat{c}_j^\dagger\big)P
\Bigg\} 
 -h\sum_jP\big(1- 2\hat{c}_j^\dagger\hat{c}_j\big)P\,.
\end{align}
By substituting $P$, a more familiar expression can be found
\begin{align}\label{eq:proj1}
P\hat{H}P=&\,-2\sum_{l<j}V_{|l-j|}\Bigg\{\hat{c}_l^\dagger|0\rangle\langle 0 |\hat{c}_j
\Bigg\} +h\sum_j\hat{c}^{\dagger}_{l}|0\rangle \langle 0|\hat{c}_{l}\,.
\end{align}
Our projected Hilbert space is composed only of the single fermion states ${|n\rangle = \hat{c}^\dagger_n|0\rangle}$ and the projector
${P=\sum_{n}\hat{c}^{\dagger}_n|0\rangle \langle 0 |\hat{c}_n}$ corresponds to the identity on this reduced Hilbert space.

\section{Spectrum of Gaussian Random Matrix.}\label{sec:dos_gaussian}
The eigenvalue spectrum of a real symmetric $L\times L$ matrix $\mathbf{M}$ with eigenvalues $M_{\lambda}$ is,
\begin{equation}\label{eq:DOS}
\nu(\lambda)=\frac{1}{L}\sum_{\lambda}\delta(\lambda-M_\lambda).
\end{equation}
Which can be redefined by analytically continuing $\lambda$ such that it contains a negative imaginary part $-i\epsilon$ to obtain,
\begin{equation}\label{eq:DOS_2}
    \nu(\lambda)=\frac{1}{\pi N}\text{Im}\left(\nu-i\epsilon-M_{\lambda})\right)^{-1}
\end{equation}
using
\begin{equation}
    \text{det}\left(\identity\lambda-\mathbf{M}\right)=\prod_{\lambda}\left(\lambda-M_{\lambda}\right)
\end{equation}
we can rewrite Eq.\,\eqref{eq:DOS} as
\begin{equation}\label{eq:DOS_LOG_DET}
    \nu(\lambda)=\frac{1}{L\pi}\text{Im}\frac{\partial}{\partial \lambda}\ln\left(\text{det}\left(\identity(\lambda-i\epsilon)-\mathbf{M}\right)\right).
\end{equation}
The logarithm can be rewritten using the following trick
\begin{equation}\label{eq:log_trick}
    \ln(x)=\lim_{n\rightarrow 0} \frac{x^n-1}{n},
\end{equation}
and by plugging this back into Eq.\,\eqref{eq:DOS_LOG_DET} we get
\begin{equation}
    \nu(\lambda)=\frac{-2}{L\pi}\text{Im}\frac{\partial}{\partial \lambda}\lim_{n\rightarrow 0}\frac{1}{n}\left(\text{det}^{-1/2}\left(\identity(\lambda-i\epsilon)-\mathbf{M}\right)^n-1\right).
\end{equation}
For later convenience, we will express the determinant as a Gaussian integral
\begin{equation}
    \text{det}^{-1/2}\left(\identity\lambda-\mathbf{M}\right)=\left(\frac{e^{i\pi/4}}{\pi^{1/2}}\right)^L\int_{-\infty}^{\infty}\prod_{i}dx_i \exp\left(-i\sum_{ij=1}^{L}x_i\left(\delta_{ij} \lambda -M_{ij}\right)x_j\right),
\end{equation}
where the imaginary part of $\lambda$ has been reabsorbed into the definition of $\lambda$ itself and it is needed to ensure convergence. We then obtain the basic result from \cite{edwards1976theigenvalue}.
\begin{equation}\label{eq:DOS}
     \nu(\lambda)=\frac{2}{L\pi}\text{Im}\frac{\partial}{\partial \lambda}\lim_{n\rightarrow 0}\frac{1}{n}\left(\left(\frac{e^{i\pi/4}}{\pi^{1/2}}\right)^{nL}\int_{-\infty}^{\infty}\prod_{i}dx_i \exp\left(-i\sum_{ij:\beta}^{L}x^{\beta}_i\left(\delta_{ij} \lambda -M_{ij}\right)x^{\beta}_j\right)-1\right).
\end{equation}
Where the integration now runs over $Nn$ variables $x^{\beta}_i$.
\subsection{Long-range mean values.}
Now, we will allow each value $M_{ij}$ to fluctuate around a mean-value which will depend on $i,j$, $M_0(|i-j|)$, such that the probability distribution of $M_{ij}$i s given by
\begin{equation}\label{eq:Prob_distribution}
    P[M_{ij}]=\frac{1}{\sqrt{2\pi \sigma^2}}\exp\left(-\frac{\left(M_{ij}-M_{0}(|i-j|)\right)^2}{2\sigma^2}\right), \hspace{1cm} M_0(|i-j|)=\frac{M_\alpha}{\left|i-j\right|^{\alpha}}.
\end{equation}
This would correspond to a disordered model which, in average, behaves in a long-range manner.

By ensemble averaging Eq.\,\eqref{eq:DOS} over all configurations $M_{ij}$ we obtain the average density of state $\rho(\lambda)$
\begin{equation}
    \rho(\lambda)=\int \prod dM_{ij} P[M_{ij}] \nu\left(\lambda; \{M_{ij}\}\right).
\end{equation}
For the Guassian probabily densities of Eq.\,\eqref{eq:Prob_distribution} we can carry out the integration over the ensemble
\begin{equation}\label{eq:DOS_M_integrated}
    \begin{split}
        \rho(\nu)= \frac{-2}{\pi L}\text{Im}\frac{\partial}{\partial \lambda} \lim_{n\rightarrow0}\frac{1}{n}\Bigg(&\left(\frac{e^{i\pi/4}}{\pi^{1/2}}\right)^{Ln}\int_{-\infty}^{\infty}\prod dx_{i}\exp\left(-i\sum_{ij,\beta}x^{\beta}_{i}\left(\lambda\delta_{ij}-\frac{M_\alpha}{|i-j|^{\alpha}}\right)x^{\beta}_{j}\right)
      \\
      &\exp\left(-\sigma^2\sum_{i,j}\left(\sum_{\beta}x^{\beta}_{i}x^{\beta}_{j}\right)^2\right)\exp\left(\sigma^2
    \sum_{i}\left(\sum_{\beta}\left(x^{\beta}_{i}\right)^2\right)^2\right)-1\Bigg)
    \end{split}
\end{equation}
Ultimately, we are interesting in the limit of $L\rightarrow \infty$ so we should look for leading order in $L$ and at most linear in $n$. Taking this into account and the $\sigma^{2}=J^{2}/L$, looking back at the exponential in Eq.\,\eqref{eq:DOS_M_integrated} we can cancel terms of second order in $n$,
\begin{equation}\label{eq:Order_n}
    \frac{J^2}{L}\sum_{ij}\left(\sum_\beta x^{\beta}_i x^{\beta}_j \right)^2=\frac{J^2}{L}\sum_{\beta}\left(\sum_i \left(x^{\beta}_i\right)^2\right)^2+\frac{J^2}{L}\sum_{\gamma\neq\beta}x^{\gamma}_i x^{\gamma}_j x^{\beta}_i x^{\beta}_j,
\end{equation}
doing so, we will only need to retain the first term in Eq.\,\eqref{eq:Order_n} where $\alpha =\beta$.
\begin{equation}\label{eq:DOS_M_integrated}
\begin{split}
        \rho(\nu)= \frac{-2}{\pi L}\text{Im}\frac{\partial}{\partial \lambda} \lim_{n\rightarrow0}\frac{1}{n}\Bigg(\left(\frac{e^{i\pi/4}}{\pi^{1/2}}\right)^{Ln}\int_{-\infty}^{\infty}\prod dx_{i}\exp\left(-i\sum_{ij}x_{i}\left(\lambda\delta_{ij}-\frac{M_\alpha}{|i-j|^{\alpha}}\right)x_{j}-\frac{J^2}{L}\left(\sum_i x^2_i\right)^2\right)^n
        \\
        -1\Bigg).
\end{split}
\end{equation}
By introducing an auxiliary field $s$, we can parametrize the second exponential in the integral above as
\begin{equation}
    \exp\left(-\frac{J^2}{L}\left(\sum_i x^2_i\right)^2\right)= \left(\frac{L}{2\pi}\right)^{1/2}\frac{\lambda}{\left(2J^2\right)^{1/2}}\int_{-\infty}^{\infty}ds\exp\left(-\frac{\lambda^2}{4J^2}Ls^2-i\lambda s \sum_i x_i^2\right),
\end{equation}
and with this the integral from Eq.\ref{eq:DOS_M_integrated} simplifies to
\begin{equation}\label{eq:DOS_M_integrated_1}
    \begin{split}
        \rho(\nu)= \frac{-2}{\pi L}\text{Im}\frac{\partial}{\partial \lambda} \lim_{n\rightarrow0}\frac{1}{n}\Bigg(&\left(\frac{e^{i\pi/4}}{\pi^{1/2}}\right)^{Ln}\left(\frac{L}{2\pi}\right)^{n/2}\frac{\lambda^n}{\left(2J^2\right)^{n/2}}
        \\
        &\int_{-\infty}^{\infty}ds\int_{-\infty}^{\infty}\prod dx_{i}\exp\left(-i\sum_{ij}x_{i}\left(\lambda(1+s)\delta_{ij}-\frac{M_\alpha}{|i-j|^{\alpha}}\right)x_{j} - \frac{\lambda^2}{4J^2}Ls^2\right)^n-1\Bigg)
    \end{split}
\end{equation}
By looking at the case of particles with long-range hopping as in \cite{defenu2021metastability}, we know that a unitary transformation $U$ exist that diagonalizes the matrix  with entries $M_{ij}=\lambda(1+s)\delta_{ij}-M_\alpha/|i-j|^{\alpha}$. It is convenient to rewrite  Eq.\,\eqref{eq:DOS_M_integrated_1} in such basis
\begin{equation}\label{eq:DOS_M_integrated_2}
    \begin{split}
        \rho(\nu)= \frac{-2}{\pi L}\text{Im}\frac{\partial}{\partial \lambda} \lim_{n\rightarrow0}\frac{1}{n}\Bigg(&\left(\frac{e^{i\pi/4}}{\pi^{1/2}}\right)^{Ln}\left(\frac{L}{2\pi}\right)^{n/2}\frac{\lambda^n}{\left(2J^2\right)^{n/2}}
        \\
        &\int_{-\infty}^{\infty}ds\int_{-\infty}^{\infty}\prod_k db_{k}\exp\left(-i\sum_{k}b_{k}\left(\lambda(1+s)-M_\alpha\epsilon_k\right)b_{k} - \frac{\lambda^2}{4J^2}Ls^2\right)^n-1\Bigg),
    \end{split}
\end{equation}
where now $\epsilon_k$ is the dispersion relation of a tight-binding model with long-range hopping parameters as in \cite{defenu2021metastability}.
\begin{align}
\epsilon_k=M_{\alpha}\frac{\text{Re}\left[\text{Li}\left(e^{ik}\right)\right]}{2}, \hspace{0.5cm} \hspace{1cm} \epsilon_k=M_{\alpha}L^{1-\alpha}\int_{0}^{1/2}\frac{\cos\left(2\pi n s\right)}{s^{\alpha}} \hspace{0.5cm} k=\frac{2\pi n}{L} \hspace{0.5cm} -\frac{L}{2}< n \leq \frac{L}{2}
\end{align}
The integral over the $b_k$ variables can be evaluated by noticing that the
\begin{equation}\label{eq:DOS_M_integrated_3}
    \begin{split}
        \rho(\nu)= \frac{-2}{\pi L}\text{Im}\frac{\partial}{\partial \lambda} \lim_{n\rightarrow0}\frac{1}{n}\Bigg(&\Bigg[\left(\frac{e^{i\pi/4}}{\pi^{1/2}}\right)^{L}\left(\frac{L}{2\pi}\right)^{1/2}\left(\frac{\pi^L}{2J^2}\right)^{1/2}\lambda\exp\left(-\frac{L}{2}\ln \lambda \right)
        \\
        &\int_{-\infty}^{\infty}ds \exp\left(-Lg(s)\right) \prod_k \left(\frac{1+s}{i(s_k-s)}\right)^{1/2} \Bigg]^n-1\Bigg)
    \end{split}
\end{equation}
where
\begin{equation}
    g(s)=\frac{\lambda^2 s^2}{4J^2}+\frac{1}{2}\ln\left(i(1+s)\right) \hspace{1cm} s_k\equiv -1 + \frac{M_\alpha \epsilon_k}{\lambda}
\end{equation}
In the limit of $L \rightarrow \infty$ can evaluate the integral in Eq.\,\eqref{eq:DOS_M_integrated_3} by using a saddle-point approximation due to exponential factor, we just have to take special care close to each $s_k$. As pointed out in \cite{edwards1976theigenvalue}, the saddle point approximation is responsible for the semicircular density of states for $|\lambda|<2J^2$. A Taylor expansion of 
$g(s)$ around $s_k$ gives
\begin{equation}
    g(s)\approx g(s_k)+\frac{\lambda}{2J^2}(\lambda_k-\lambda)(s-s_k)+\frac{\lambda^2}{4}\left(\frac{1}{J^2}-\frac{1}{M_\alpha^2\epsilon_k^2} \right)(s-s_k)^2+...
\end{equation}
with
\begin{equation}
    \lambda_k=M_\alpha\epsilon_k+\frac{J^2}{M_\alpha\epsilon_k}
\end{equation}
therefore, when $\lambda=\lambda_n$, $g(s)$ has a turning point. One can also see that for for $M_0\epsilon_n^2>J$ this turning point is a minimum, while it is a maximum for $M_0\epsilon_n^2<J$. Moreover this turning points lie outside of the semicircular continuum of eigenvalues. The integral in Eq.\,\eqref{eq:DOS_M_integrated_3} is evaluated by retaining the first two terms of the Taylor expansion of $g(s)$. The result is
\begin{equation}\label{eq:DOS_M_integrated_4}
    \begin{split}
        \rho(\nu)= \frac{-2}{\pi L}\text{Im}\frac{\partial}{\partial \lambda} \lim_{n\rightarrow0}\frac{1}{n}\Bigg(&\Bigg[\left(\frac{e^{i\pi/4}}{\pi^{1/2}}\right)^{L}\left(\frac{L}{2\pi}\right)^{1/2}\left(\frac{\pi^L}{2J^2}\right)^{1/2}\exp\left(-\frac{L}{2}\ln \lambda \right)
        \\
        & \hspace{2cm}\exp\left(-Lg(s_k)\right)\exp\left(-\frac{1}{2}\ln\left(\lambda-\lambda_k\right)\right)\Bigg]^n-1\Bigg)
    \end{split}
\end{equation}
with
\begin{equation}
    g(s_k)=-\frac{\lambda^2}{4J^2}\left(\frac{M_\alpha\epsilon_k}{\lambda}-1\right)^2+\frac{\ln \lambda}{2}-\frac{\ln M_\alpha\epsilon_k}{2}-\frac{i\pi}{4}
\end{equation}
Plugging this back into Eq.\,\eqref{eq:DOS_M_integrated} and using Eq.\,\eqref{eq:log_trick} the eigenvalue spectrum for $M_\alpha \epsilon_k > J$ is
\begin{equation}
    \rho(\lambda)=-\frac{2}{L\pi}\text{Im}\frac{\partial}{\partial\lambda}\left[-\frac{1}{2}\ln\left(\lambda-\lambda_k\right)\right]=\frac{1}{L}\delta(\lambda-\lambda_k)
\end{equation}
The eigenvalue spectrum of a large random symmetric matrix where the mean value has a power-law dependence with the distance from the diagonal is
\begin{equation}
    \rho(\lambda)=\frac{\sqrt{4J^2-\lambda^2}}{2\pi J^2}+\frac{1}{N}\sum_{k\forall k\in \epsilon_k<\epsilon^*}\delta(\lambda-\lambda_k) \hspace{0.5cm} \text{where}\hspace{0.5cm} \epsilon^*\equiv \frac{J}{M_\alpha}.
\end{equation}
We would like to point out that this computation assumes that we can analyze $g(s)$ in the neighborhood of each $s_k$ independently, which is valid when our energy levels $e_k$ are gaped. Due to this fact we can only predict correctly the density of states of strongly interacting systems $\alpha<d$ where we know the spectrum remains gaped in the thermodynamic limit.
\begin{figure}
    \centering
\includegraphics[width=.59\linewidth,trim={0 0 0 0},clip]{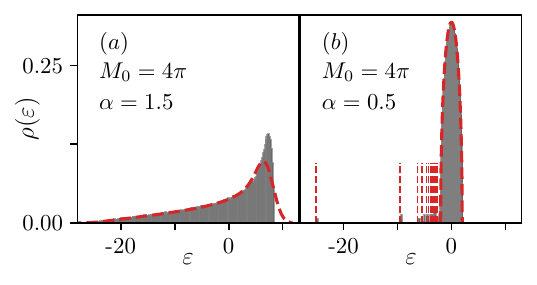}
    \caption{Panels~(a)-(b): Density of states (DOS) of a real symmetric random matrix with mean-long range power-law decaying off diagonal terms and gaussian disorder. Such DOS corresponds to the DOS of a free-particle with off-diagonal disorder on top of power-law decaying long-range hoppings amplitudes. Note how the appearance of low-energy gapped states only happens for $\alpha<1$.}
    \label{Fig1}
\end{figure}
\section{Dynamical Mean-Field Theory.}\label{sec:DMFT}
Dynamical mean-field theory replaces a lattice many-body problem to a single-site quantum impurity one. By doing so, one does not consider spatial fluctuations but only keeps track of temporal ones. For a Hubbard model with on site interactions the impurity effective dynamics are described by the imaginary-time action
\begin{equation}
    S_{\text{eff}}=-\int_{0}^{\beta}d\tau\int_{0}^{\beta}d\tau' \sum_{\sigma}c^{\dagger}_{0,\sigma}(\tau)\mathscr{G}^{-1}_{0}(\tau-\tau')c_{0\sigma}(\tau')-iU\int_{0}^{\beta}d\tau n_{0\uparrow}(\tau)n_{0\downarrow}(\tau)
\end{equation}
where $(c_{0\sigma},c^{\dagger}_{0\sigma})$ are the fermionic degrees of freedom of the so-called impurity.

Within DMFT the self-consistent equation that determines the relation of the effective single site Green's function and the on-site lattice interacting Green's function is given by
\begin{equation}\label{eq:weiss_self}
    \mathscr{G}^{-1}_0(i\omega_n)=i\omega_n+\mu+G^{-1}(i\omega_n)-R[G],
\end{equation}
where $R[G]$ is the reciprocal function defined as
\begin{equation}\label{eq:reciprocal}
    \Tilde{\rho}(\xi)\equiv \int d\varepsilon\frac{\rho(\varepsilon)}{\xi-\varepsilon}, \hspace{1cm} R[\Tilde{\rho}(\xi)]=\xi.
\end{equation}
In general, one cannot find an explicit connection between $\mathscr{G}^{-1}_0(i\omega_n)$ and $G(i\omega_n)$.

\subsection{Long-range hopping.}
As a pedagogical example we will look at the case where the particles are allowed to hop everywhere in the lattice. The density of states of such systems consist of an isolated eigenvalue plus an accumulation point at zero energy
\begin{equation}
    \rho(\varepsilon)=\frac{1}{L}\delta(\varepsilon-\varepsilon_0)+\frac{L-1}{L}\delta(\varepsilon)
\end{equation}
for such DOS after computing the reciprocal function Eq.\ref{eq:reciprocal}, the self-consistent relation between the Weiss field and the local Green's function reads as
\begin{equation}\label{eq:weiss_self}
    \mathscr{G}^{-1}_0(i\omega_n)=i\omega_n+\mu+G^{-1}(i\omega_n)-\frac{\left( L+\varepsilon_0+\varepsilon_0 LG(i\omega_n)\right) - \left(\left(\varepsilon_0-L+\varepsilon_0 LG(i\omega_n)\right)^2+4\varepsilon_0 L\right)^{1/2} }{ 2 L G(i\omega_n) },
\end{equation}
which, on the limit of $L\rightarrow \infty$ becomes
\begin{equation}
    \mathscr{G}^{-1}_0(i\omega_n)=i\omega_n+\mu-\frac{1}{L}\frac{\varepsilon_0}{\varepsilon_0 G(i\omega_n) -1}=i\omega_n+\mu-\mathcal{O}(1/L).
\end{equation}
In the limit of larger $N$ one recovers the atomic limit where all sites are disconnected and particles cannot hop in between them.

The situation changes slightly when the hopping are not rescaled under Kac's prescription. Substituing $\varepsilon_0\equiv L \varepsilon_0$ and expanding Eq.~\eqref{eq:weiss_self}

\begin{equation}
\mathscr{G}^{-1}_0(i\omega_n)=i\omega_n+\mu-\frac{1}{4L}\frac{\left(\varepsilon_0+1\right)^2}{\varepsilon_0 G^2(i\omega_n)}=i\omega_n+\mu-\mathcal{O}(1/L).
\end{equation}

\subsection{Semi-circular density of states.}
For the case of a semi-circular density of states of the form,
\begin{equation}
    \rho(\varepsilon)=\frac{1}{2\pi J^2}\sqrt{4J^2-\varepsilon^2}
\end{equation}
one obtains
\begin{equation}
    \Tilde{\rho}(\xi)=\frac{\xi-\sqrt{\xi^2-4J^2}}{2J^2}, \hspace{1cm} R[\Tilde{\rho}(G)]=J^2G+\frac{1}{G}.
\end{equation}
which leads an explicit form between the local Green's function and the effective Weiss field
\begin{equation}
    \mathscr{G}^{-1}_0(i\omega_n)=i\omega_n+\mu-J^2 G(i\omega_n).
\end{equation}
For our long-range interacting systems with disorder, we have seen how the density of states consist of a finite set of discrete energy levels $\varepsilon_{n}\in $ on top of a semi-circular continuum. This leads to the following Hilbert transform,
\begin{equation}
    \tilde{D}(\xi)=\frac{1}{L}\frac{\xi-\sqrt{\xi^2-4J^2}}{2J^2}+\sum_{\varepsilon_n \in {\varepsilon*}}\frac{1}{L}\frac{1}{\xi-\varepsilon_n},
\end{equation}
an explicit expression for the reciprocral function cannot be found but it is easy to see that on the limit of $L\rightarrow \infty$ the set of discrete points, with a factor of $1/L$ in front, can be thought as a correction tot the general semi-circular form,
\begin{equation}
    \mathscr{G}^{-1}_0(i\omega_n)=i\omega_n+\mu-t^2 G(i\omega_n)+\mathcal{O}(1/L).
\end{equation}

\section{Fidelity computation.}\label{sec:fidelity_supp}
The fidelity consist on the overlap between the same state evolved under different Hamiltonians. In the case of our interest, we would like to compare our random hopping Hamiltonian with long-range tails to one where the hoping amplitudes are sampled from a Gaussian distribution zero mean. 
\begin{equation}\label{eq:Hamiltonians}
    \begin{split}
&\hat{H}_0=\sum_{ij}t_{ij}\left(\hat{c}^{\dagger}_{i,\uparrow}\hat{c}_{j,\uparrow}+\hat{c}^{\dagger}_{i,\downarrow}\hat{c}_{j,\downarrow} \right)+U\sum_i\hat{c}^{\dagger}_{i,\uparrow}\hat{c}_{i,\uparrow}\hat{c}^{\dagger}_{i,\downarrow}\hat{c}_{i,\downarrow},
\\
&\hat{H}_\alpha=\sum_{ij}M_{ij}\left(\hat{c}^{\dagger}_{i,\uparrow}\hat{c}_{j,\uparrow}+\hat{c}^{\dagger}_{i,\downarrow}\hat{c}_{j,\downarrow} \right)+U\sum_i\hat{c}^{\dagger}_{i,\uparrow}\hat{c}_{i,\uparrow}\hat{c}^{\dagger}_{i,\downarrow}\hat{c}_{i,\downarrow}.
    \end{split}
\end{equation}
As an initial state we take the ground state of the kinetic term in $H_0$ at half-filling, which we denote as $\ket{F.S.}$.
\begin{equation}\label{eq:fidelity}
    F(t)=\left|\bra{\text{F.S.}} e^{it\hat{H}_0}e^{-it\hat{H}_\alpha}\ket{\text{F.S.}}\right|^2.
\end{equation}
In order to be able to capture correctly the physics behind the long-range nature of the tails such that they are not washed away by the randomness of the system a few assumptions are in order. Since the eigenvectors of a Guassian $L\times L$ orthogonal matrix are random unit vectors on the surface of a $L$ dimensional sphere, a naive computation of the fidelity for two different realizations of disorder would quickly tend to zero due to a orthogonality catastrophe effect. We will, therefore, assume that the eigenvectors between both realizations are the same and that only their eigen-energies change in the presence of long-range tails.

Both Hamiltonians in Eq.\ref{eq:Hamiltonians} can be rewritten in terms of the random vectors that diagonalize the hoping term as,
\begin{equation}
    \begin{split}
\hat{H}_0=\sum_{l}\varepsilon_n\left(\hat{b}^{\dagger}_{n,\uparrow}\hat{b}_{n,\uparrow}+\hat{b}^{\dagger}_{n,\downarrow}\hat{b}_{n,\downarrow} \right)+U\sum_i\sum_{l_1,l_2,l_3,l_4}a^*_{i,l_1}a_{i,l_2}a^*_{i,l_3}a_{i,l_4}\hat{b}^{\dagger}_{l_1,\uparrow}\hat{b}_{l_2,\uparrow}\hat{b}^{\dagger}_{l_3,\downarrow}\hat{b}_{l_4,\downarrow},
        \\
\hat{H}_\alpha=\sum_{l}\varepsilon'_n\left(\hat{b}^{\dagger}_{n,\uparrow}\hat{b}_{n,\uparrow}+\hat{b}^{\dagger}_{n,\downarrow}\hat{b}_{n,\downarrow} \right)+U\sum_i\sum_{l_1,l_2,l_3,l_4}a^*_{i,l_1}a_{i,l_2}a^*_{i,l_3}a_{i,l_4}\hat{b}^{\dagger}_{l_1,\uparrow}\hat{b}_{l_2,\uparrow}\hat{b}^{\dagger}_{l_3,\downarrow}\hat{b}_{l_4,\downarrow}.
    \end{split}
\end{equation}
The coefficients $a_{i,l}$ are random numbers restricted to fulfill $\sum_{l}|a_{i,l}|^2=1$ and $\varepsilon_n$,$\varepsilon'_n$ differ only in the finite subset of energy levels $n$ that lie outside the semicircular.

Again, the interacting term is treated as a small perturbation and expanded up to second order using unitary perturbation theory. We start with a Hamiltonian of the form $\hat{H}=\hat{H}_0+g\hat{H}_{1}$ and tranform it via a canonical transformation $e^{\hat{S}}$ chosen to cancel, at each order of $g$, every off-diagonal term.
\begin{equation}
\begin{split}
    &\hat{S}=g\hat{S}_1+\frac{1}{2}g^2\hat{S}_2+\mathcal{O}(g^3),
    \\
    &\hat{H}_{\text{diag}}=e^{\hat{S}}\hat{H}e^{-\hat{S}}=H_0+g\left(\hat{H}_1+[\hat{S}_1,\hat{H}_0]\right)+g^2\left(\frac{1}{2}[\hat{S}_2,\hat{H}_0]+[\hat{S}_1,\hat{H}_1]+\frac{1}{2}[\hat{S}_1,[\hat{S}_1,\hat{H}_0]]\right)+\mathcal{O}(g^3)
\end{split}
\end{equation}

the transformed Hamiltonian and the unitary transformation read

\begin{equation}
\begin{split}
\hat{H}_{\text{diag}} &= e^{\hat{S}}\hat{H} e^{-\hat{S}} = \hat{H}_0 + g\hat{H}_{\text{diag}}^{(1)} + g^2\hat{H}_{\text{diag}}^{(2)} + O(g^3), \\
\bra{n}\hat{S}_1\ket{m} &= \begin{cases}
\bra{n}\hat{H}_1\ket{m} & \text{if } n \neq m \\
0 & \text{if } n = m
\end{cases}, \\
\bra{n}\hat{S}_2\ket{m} &= \begin{cases}
\bra{n}[\hat{S}_1,\hat{H}_1 + \hat{H}_{\text{diag}}^{(1)}]\ket{m} & \text{if } n \neq m \\
0 & \text{if } n = m
\end{cases}, \\
\hat{H}_{\text{diag}}^{(i)} &= \sum_{n} \ket{n}E^{(i)}_n\bra{n}, \\
E^{(1)}_n &= \bra{n}\hat{H}_1\ket{n}, \quad E^{(2)}_n = \sum_{m\neq n}\frac{|\bra{n}\hat{H}_1\ket{m}|^2}{E_n - E_m}
\end{split}
\end{equation}
the operator $\hat{S}$ which transforms the Hamiltonian $H_0$ is, to first order
\begin{equation}
    S=U\sum_i\sum'_{l_1,l_2,l_3,l_4}\frac{a^*_{i,l_1}a_{i,l_2}a^*_{i,l_3}a_{i,l_4}}{\varepsilon_{l_1}-\varepsilon_{l_2}+\varepsilon_{l_3}-\varepsilon_{l_4}}\hat{b}^{\dagger}_{l_1,\uparrow}\hat{b}_{l_2,\uparrow}\hat{b}^{\dagger}_{l_3,\downarrow}\hat{b}_{l_4,\downarrow}
\end{equation}
where the $\sum'$ denotes that the term $l_1=l_2 \land l_3=l_4$ is excluded. It is straightforward to see that the $S'$ which transform $H_{\alpha}$ has the same form as $S$ but with the substitution of $\varepsilon_i \rightarrow \varepsilon'_i$.

With this, we can go back to Eq.\ref{eq:fidelity} and express it in terms of $S$ and $S'$.
\begin{equation}
    F(t)=\left|\bra{\text{F.S.}}_0 e^{S(0,0)-S(0,-t)}e^{S'(-t,0)-S'(0,0)}\ket{\text{F.S.}}_0\right|^2
\end{equation}
 where we have defined $S(-t,0)=e^{-it\hat{H}_{0\text{diag}}}Se^{it\hat{H}_{0\text{diag}}}$ and $S(0,-t)=e^{-it\hat{H}_{\alpha\text{diag}}}Se^{it\hat{H}_{\alpha\text{diag}}}$ and already canceled any phase factors given by the evaluation of diagonal terms $e^{-it\hat{H}_{\alpha\text{diag}}}\ket{F.S.}$ since the fidelity is defined as the modulus of the overlap.

 We then expand the exponentials up to second order and evaluate to obtain
\begin{equation}
    \begin{split}
        F(t)=&1- 8U^2\sum_{ij}\sum_{\substack{l_1, l_2\\ l_3\neq l_3}}|a_{i,l_1}|^2|a_{j,l_2}|^2 a^*_{i,l_3}a_{j,l_3}a^*_{j,l_4}a_{i,l_4}\left(n_{l_1}-\delta_{l_1,l_3}(n_{l_1}-1))n_{l_3}n_{l_3}(1-n_{l_4})\right)
        \\
        &\hspace{1cm}\times \Bigg[\frac{\sin^2\left(\left(\varepsilon_{l_3}-\varepsilon_{l_4}\right)/2\right)}{\left(\varepsilon_{l_3}-\varepsilon_{l_4}\right)^2}+\frac{\sin^2\left(\left(\varepsilon'_{l_3}-\varepsilon'_{l_4}\right)/2\right)}{\left(\varepsilon'_{l_3}-\varepsilon'_{l_4}\right)^2}
        \\
&\hspace{2cm}-2\cos\left(\frac{\varepsilon_{l_3}-\varepsilon_{l_4}-\varepsilon'_{l_3}+\varepsilon'_{l_4}}{2}\right)\frac{\sin\left(\left(\varepsilon_{l_3}-\varepsilon_{l_4}\right)/2\right)\sin\left(\left(\varepsilon'_{l_3}-\varepsilon'_{l_4}\right)/2\right)}{\left(\varepsilon_{l_3}-\varepsilon_{l_4}\right)\left(\varepsilon_{l'_3}-\varepsilon_{l'_4}\right)}\Bigg]
        \\
        &-U^2\sum_{ij}\sum_{\substack{l_1\neq l'_1 l_3\neq l'_3}}a^*_{i,l_1}a_{j,l_1}a^*_{j,l'_1}a_{i,l'_1} a^*_{i,l_3}a_{j,l_3}a^*_{j,l_0}a_{i,l_0}\left(n_{l_1}n_{l_3}(1-n_{l'_1})(1-n_{l_0})\right)
        \\
        &\hspace{1cm}\times \Bigg[\frac{\sin^2\left(\left(\varepsilon_{l_1}-\varepsilon_{l_2}+\varepsilon_{l_3}-\varepsilon_{l_4}\right)/2\right)}{\left(\varepsilon_{l_1}-\varepsilon_{l_2}+\varepsilon_{l_3}-\varepsilon_{l_4}\right)^2}+\frac{\sin^2\left(\left(\varepsilon'_{l_1}-\varepsilon'_{l_2}+\varepsilon'_{l_3}-\varepsilon'_{l_4}\right)/2\right)}{\left(\varepsilon'_{l_1}-\varepsilon'_{l_2}+\varepsilon'_{l_3}-\varepsilon'_{l_4}\right)^2}
        \\
&\hspace{2cm}-2\cos\left(\frac{\varepsilon_{l_1}-\varepsilon_{l_2}+\varepsilon_{l_3}-\varepsilon_{l_4}-\varepsilon'_{l_1}+\varepsilon'_{l_2}-\varepsilon'_{l_3}+\varepsilon_{l_4}}{2}\right)
\\
&\hspace{4cm}\times\frac{\sin\left(\left(\varepsilon_{l_1}-\varepsilon_{l_2}+\varepsilon_{l_3}-\varepsilon_{l_4}\right)/2\right)\sin\left(\left(\varepsilon'_{l_1}-\varepsilon'_{l_2}+\varepsilon'_{l_3}-\varepsilon'_{l_4}\right)/2\right)}{\left(\varepsilon_{l_1}-\varepsilon_{l_2}+\varepsilon_{l_3}-\varepsilon_{l_4}\right)\left(\varepsilon_{l'_1}-\varepsilon_{l'_2}+\varepsilon_{l'_3}-\varepsilon_{l'_4}\right)}\Bigg].
    \end{split}
\end{equation}
As a last step we substitute the sum over the sites $\{i,j\}$ by averages over disorder. To do so notice that $a_{i,l}$ are components of a random unit vectors and perform the average by using the following probability distribution,
\begin{equation}
    P[|a_{i,j}|^2]=\frac{\Gamma\left(\frac{L}{2}\right)}{\sqrt{\pi}\Gamma\left(\frac{l-1}{2}\right)}\frac{\left(1-|a_{i,j}|^2\right)^{(L-3)/2}}{|a_{i,j}|}, \hspace{1cm} \lim_{L\rightarrow \infty}P[z=y /L]=\frac{e^{-z/2}}{\sqrt{2\pi z}}
\end{equation}

\begin{figure}[h]
    \centering
\includegraphics[width=.59\linewidth,trim={0.1cm 0.12cm 0.02cm 0.02cm},clip]{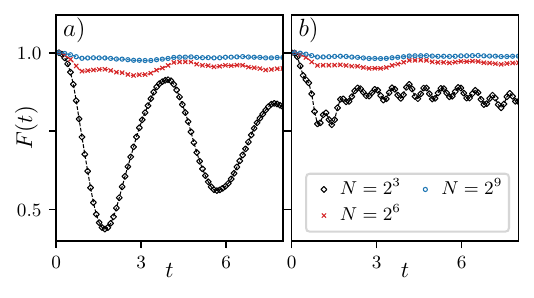}
    \caption{Fidelity decay when the system is evolved under purely random Gaussian hopping, $H_{\rm Bethe}$ (panel (a))  vs one with random gaussian hoppings with avergae power-law decaying hooping amplitudes as described by Eq.\,\eqref{eq:Prob_distribution}. For panel (b): $\alpha=0.5$ and $H_\alpha$. In both cases $U =1$, $\sigma=1$ and $M_0=-2\pi$. The initial state corresponds to a filled Fermi-sea at half filling of the purely disorder case $H_0$. Three different systems sizes are shown $L=2^{3},2^{6},2^{9}$ (black, red, blue). The smallest system shows a dependence on the discrete low-energy states which is washed away as the system size increases.}
    \label{fig:Fidelity}
\end{figure}


    
    
    

\end{document}